\documentclass{aa} 
\usepackage{graphicx}
\newcommand{\excs}{\extracolsep{\fill}} 
\begin{document}
   \title{Adaptive optics imaging survey of the Tucana-Horologium association}
    \subtitle{}
\titlerunning{}
\authorrunning{Chauvin et al.}
\author{
        G. Chauvin\inst{1},
        M. Thomson\inst{1}\inst{,2},
        C. Dumas\inst{3},
        J-L. Beuzit\inst{1},
        P. Lowrance\inst{3},
        T. Fusco\inst{4},
        A-M. Lagrange\inst{1},
        B. Zuckerman\inst{5}
 \and
        D. Mouillet\inst{6}
}
\offprints{Ga\"el Chauvin, \email{gchauvin@obs.ujf-grenoble.fr}}
    \institute{
                $^{1}$Laboratoire d'Astrophysique, Observatoire de 
Grenoble, 414, Rue de la piscine, Saint-Martin 
d'H\`eres, France\\
                $^{2}$Imperial College of Science, Technology and 
Medecine, Exhibition  Road, London SW7 2AZ, England\\
                $^{3}$Jet Propulsion Laboratory, Mail Stop 183-501, California Institute of Technology, 4800 Oak Grove Drive, Pasadena, CA 91109\\
                $^{4}$ONERA, BP52 29 Avenue de la Dividion Leclerc 
Ch\^atillon Cedex, France \\
$^{5}$Department of Physics and Astronomy, University of California, Los Angeles, 8371 Math Science Building, Box 951562, CA 90095-1562, USA\\
                $^{6}$Laboratoire d'Astrophysique, Observatoire 
Midi-Pyr\'en\'enes, Tarbes, France\\
              }
\date{Received: 29-07-02  / Accepted: 04-03-03}
\abstract{
We present the results of an adaptive optics (AO) imaging survey of the common associations of Tucana and Horologium, carried out at the ESO 3.6m telescope with the ADONIS/SHARPII system. Based on our observations of two dozen probable association members, HIP~1910 and HIP~108422 appear to have low-mass stellar companions, while HIP~6856 and GSC~8047-0232 have possible sub-stellar candidate companions. Astrometric measurements, performed in November 2000 and October 2001, indicate that HIP~1910 B likely is bound to its primary, but are inconclusive in the case of the candidate companion to HIP~6856. Additional observations are needed to confirm the HIP~6856 companionship as well as for HIP~108422 and GSC~8047-0232.
}
%
%
\maketitle

%


\section{Introduction}
Young, nearby stellar associations are ideal targets for the direct imaging of sub-stellar companions: (i) their proximity ($<100$~pc) allows the exploration of the faint circumstellar environment at relatively small distances from the star, and (ii) sub-stellar objects are hotter and brighter when young ($\sim1$0 Myr) and therefore can be more easily detected than more evolved counterparts.

Based on data in the \textit{IRAS} and \textit{Hipparcos} catalogs, Zuckerman \& Webb (2000) identified comoving stars of the Tucana association within 50~pc from earth. They tentatively assigned an age of $\sim40~$Myr for this comoving group based on the H$\alpha$ emission lines of three members.
A consistent estimation of 10-30~Myr was found by Stelzer \& Neuh\"auser (2001) based on \textit{ROSAT} data and comparison of the X-ray luminosity functions of Tucana and star forming regions. Zuckerman et al. (2001a) extended the number of probable members to 36 and argue that the Tucana association and the Horologium association, identified from \textit{ROSAT} measurements by Torres et al.~(2000), should be considered as part of the same stream. They may form a large, gravitationally unbound, association composed of $\sim50$ stars. 

Lowrance et al. (2000)  reported the HST/NICMOS detection of a possible sub-stellar companion to the proposed Tucana member HR~7329, with a mass less than 50~M$_{\rm{Jup}}$ (Burrows et al. 1997). This exciting result, confirmed by spectroscopy and astrometry (Guenther et al. 2001), encouraged us to perform an AO imaging survey of this large association, searching for sub-stellar companions close to member stars. Ironically, HR~7329 appears not to be a member of the Tucana association, but rather of the substantially younger $\beta$ Pictoris moving group (Zuckerman et al. 2001b). In addition, HIP~92680 and HIP~95270 attributed initially to Tucana by Zuckerman \& Webb (2000) also appear rather to be $\beta$ Pictoris group members (Zuckerman et al. 2001b). But before this was understood, we observed HIP~95270 with the ADONIS/SHARPII AO system on the ESO 3.6~m telescope. 

In Sect. 2, we present the instrumental configuration and the observational strategy we adopted in both classical and coronagraphic imaging. We also give the detection performances commonly achieved during the survey.
In Sect. 3, we present our photometric results for the three visual systems HIP~108422, GSC~8047-0232 and HIP~6856 and the confirmed binary HIP~1910~AB. We also report astrometric measurements obtained on November 2000 and October 2001 for the stars HIP~6856 and HIP~1910. The probability that four detected possible secondaries are associated is discussed. In Sect. 4-7, we discuss the photometry of the three companion candidates (CCs) and HIP~1910~AB based on evolutionnary models of Baraffe et al. (2002, 1998) and Chabrier et al. (2000). Our purpose is to test if the photometric results are consistent with physically bound objects and, if so, to determine the corresponding stellar parameters. These evolutionary models provide the most accurate comparison with present observations of young objects but are not as reliable for ages of a few tens of Myr because of unknown initial conditions.


\section{Observation Strategy and performances}

\begin{figure}[b]
\begin{center}
\vfill
\includegraphics[width = 8.5cm]{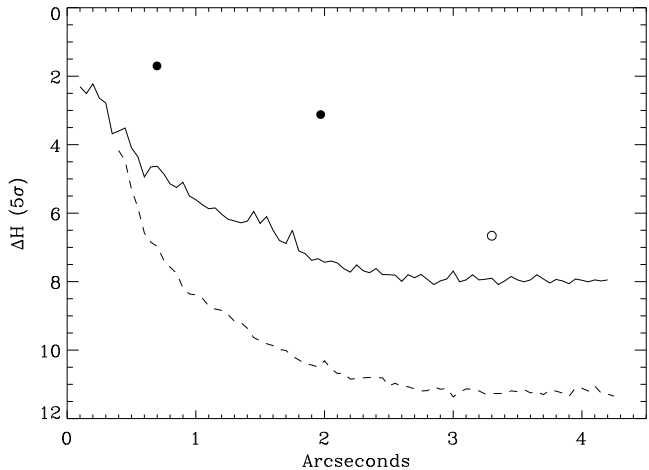}
\caption[]{\label{fig:limdet}{H band (pixel to pixel) detection limits during the observations of HIP~1113 (G6V; V = 8.8) in classical mode during 30 $\times$ 2~s (\textit{solid line}) and in coronagraphic mode during 15 $\times$ 20~s (\textit{dashed line}). Filled circles are HIP~1910~B and the CC to HIP~108422 detected in classical mode. The open circle is the CC to GSC~80047-0232 detected in coronagraphy. }}
\end{center}
\end{figure}

High contrast observations were performed with the AO ADONIS/SHARPII system over two distinct periods:  10-13 November 2000 and 28-29 October 2001. Of suggested members of the Tucana-Horologium association, 24 were observed in J, H and K bands, with and without coronagraphy, to explore the domain between $0.5~\!''$ and $6.0~\!''$ from the central star (i.e., 25~AU to 300~AU for a distance of 50~pc). In both runs, the pixel scale on the detector is estimated to be $0.0496\pm0.0003~\!''$/px. The corresponding field of view is $12.7~\!''\times12.7~\!''$. The orientation direction of true north was found to lie to the west of vertical by $0.47^o\pm0.5$ on November 2000 and $0.56^o\pm0.5$ on October 2001. The plate scale and field rotation are based on astrometric calibrations of ADONIS/SHARPII+/NICMOS3\footnote{\emph
http://www.bdl.fr/priam/adonis/} obtained from observations of the astrometric reference field ($\theta$ Ori).

For each coronagraphic observation, a reference star was observed for PSF subtraction after scaling to the brightness of the science object. The coronagraphic masks were $1.0~\!''$ in diameter on November 2000 and $0.84~\!''$ on October 2001.

Seeing conditions were relatively good during both observing runs, ranging between $0.6~\!''$ and $1~\!''$, except on 28 October 2001 where the average seeing was $1.3~\!''$. The corresponding H-band strehl and FWHM were $\sim45\%$ and $\sim0.1~\!''$ under good seeing conditions and reduced to $\sim30\%$ and $\sim0.11~\!''$ on 28 October 2001. The (pixel to pixel) detection limits at $5\sigma$ were estimated from observations with and without coronagraphy for each object. The case of HIP~1113 is treated as an example in Fig. \ref{fig:limdet} and presents the typical detection performances achieved on 29 October 2001, in terms of separation and contrast with the primary star. The CC detected near HIP~108422 and GSC~8047-0232, as well as the companion HIP~1910~B, are overplotted.

\section{Results}

\begin{figure}[b]
\begin{center}
\vfill
\includegraphics[width = 8.5cm,angle=-90]{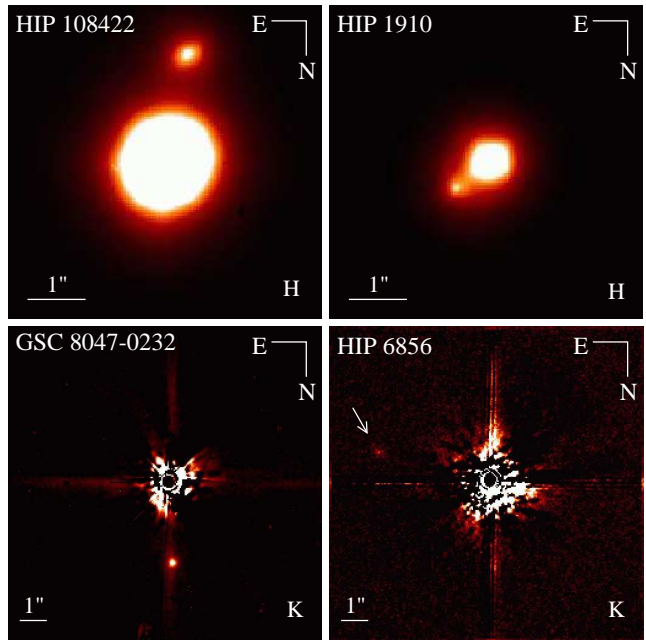}
\caption[]{ADONIS/SHARPII AO classical imaging of HIP~108422 and HIP~1910~AB obtained in H-band (linear stretch) and coronagraphic imaging of GSC 8047-0232 and HIP~6856, respectively in K band (linear stretch).}
\label{fig:ao}
\end{center}
\end{figure}

\begin{table*}[t]
\begin{minipage}[t]{16 cm}
      \caption[]{Photometry and relative positions of HIP~108422, GSC~8047-0232 and their CCs, of the CC to HIP~6856 and of HIP~1910~A and B. Distances of HIP~108422, HIP~1910~AB and HIP~6856 are obtained from the Tycho catalog (H{\o}g et al. 2000).}
         \label{tab:phot}
      \[\centering
\begin{tabular}{llllllll}
\hline
\hline
\noalign{\smallskip}
Source & UT Date                 & J    & H    & K & separation & P.A.    & distance \\
                         &    &(mag.)&(mag.)&(mag.)& ($\!''$) &($^o$)   & (pc)     \\
\noalign{\smallskip}
\hline
\noalign{\smallskip}
HIP~108422 & 2001 Oct 28           & 7.35 $\pm$ 0.07 & 6.90 $\pm$ 0.06 & 6.83 $\pm$ 0.06 & & & 54.9 $^{+3.8}_{-3.3}$   \\
HIP~108422~CC &   2001 Oct 28           & 10.67 $\pm$ 0.16 & 10.02 $\pm$ 0.15 & 9.78 $\pm$ 0.12 & 1.979 $\pm$ 0.035 &          $192.1\pm0.9$ &         \\
\noalign{\smallskip}
\hline
\noalign{\smallskip}
HIP~1910~A &     2000 Nov 12            & -    & 7.86 $\pm$ 0.09 & 7.69 $\pm$ 0.07& & &46.3 $^{+5.3}_{-4.3}$    \\
HIP~1910~B & 2000 Nov 12         & -    & 9.55 $\pm$ 0.16 & 9.22 $\pm$ 0.1& 0.698 $\pm$ 0.029     & 49.6 $\pm2.5$   &         \\
HIP~1910~B & 2001 Oct 28 & -& -& -& 0.699 $\pm$ 0.030 &50.2 $\pm2.5$ & \\
\noalign{\smallskip}
\hline
\noalign{\smallskip}
GSC~8047-0232    &  2001 Oct 29          & 9.06 $\pm$ 0.09 & 8.54 $\pm$ 0.05 & 8.45 $\pm$ 0.07 & & &(60 $\pm$ 25)    \\
GSC~8047-0232~CC &   2001 Oct 29         & 16.25 $\pm$ 0.25 & 15.2 $\pm$ 0.18 & 14.9 $\pm$ 0.2 & 3.210 $\pm$ 0.118 &  359.2 $\pm2.3$ &       \\
\noalign{\smallskip}
\hline
\noalign{\smallskip}
HIP~6856 &  2000 Nov 12              &  -   & -    & -    & & &37.1$^{+1.8}_{-1.6}$    \\
HIP~6856~CC & 2000 Nov 12 & -& -&- & 4.865 $\pm$ 0.129 & 106.5 $\pm1.7$ &\\
HIP~6856~CC & 2001 Oct 29 & 17.97 $\pm$ 0.2 &   -  & 17.4 $\pm$ 0.15 & 4.785 $\pm$ 0.129 & 106.6 $\pm1.7$   &          \\
\noalign{\smallskip}
\hline     
        \end{tabular}
      \]
\end{minipage}
\end{table*}

Twenty four suggested members of the Tucana and Horologium associations were observed; no CCs were detected around the following stars: HIP~1113, HIP~1481, HIP~1993, HIP~2729, HIP~3556, HIP 6485, HIP~9902, HIP~93815, HIP~95270, HIP~105388, HIP~107345, HIP~107649, HIP~107947, GSC 8497-0995, GSC 8056-0482, HD 13183, CoD-60416, CoD-53544 and CoD-65149. We confirmed the known binary system HR~7329~AB. Figure~\ref{fig:ao} shows images of the HIP~1910, HIP~108422, GSC~8047-0232 and HIP~6856 systems. The binary system HIP~1910~AB and the CC detected around HIP~108422 were observed without coronagraphy. HIP~1910~AB and HIP~108422 are members of the Tucana association (Zuckerman \& Webb 2000; Zuckerman et al. 2001a). Two fainter CCs were detected with coronagraphy around GSC~8047-0232 and HIP~6856, members of the Horologium association (Torres et al. 2000). 

\subsection{Photometry}

The photometry of HIP~1910~AB and of the visual systems HIP~108422, GSC~8047-0232 and HIP~6856 was calibrated with photometric standards from the \textit{Hubble Space Telescope} near-infrared standards : S055D (04:18:18.9 -69:27:35), S294D (00:33:15.2 -39:24:10), S209D (08:01:15.4 -50:19:33) and  S301D (03:26:53.9 -39:50:38). The results are reported in Table~\ref{tab:phot}. To estimate the flux and the position of each component in the visual binary HIP~108422, the deconvolution algorithm of V\'eran \& Rigaut (1998) was used. In the closest system HIP~1910~AB, we used the myopic deconvolution algorithm MISTRAL (Fusco et al. 1999; Conan et al. 2000). For the two fainter CCs around GSC~8047-0232 and HIP~6856, detected with coronagraphy, aperture photometry was performed. A small aperture of radius $1.0~\!''$ was used to avoid subtraction residuals. In the case of the CC to GSC~8047-0232, the photometric estimation was more difficult because this object was contaminated by a diffraction spike of the secondary mirror. Despite several attempts to correct our image for the spike contribution, we could not avoid increased photometric uncertainty. 

\subsection{Astrometry}

The systems HIP~1910~AB and HIP~6856 were observed on November 2000 and October 2001. The corresponding astrometry could be determined for both systems. We took into account errors coming from the relative position estimation of both components in each system, the instrumental uncertainties (pixel scale and orientation error; see Sect. 2), as well as the proper motions errors given by the Tycho catalog (H{\o}g et al. 2000). 

\begin{figure*}[t]
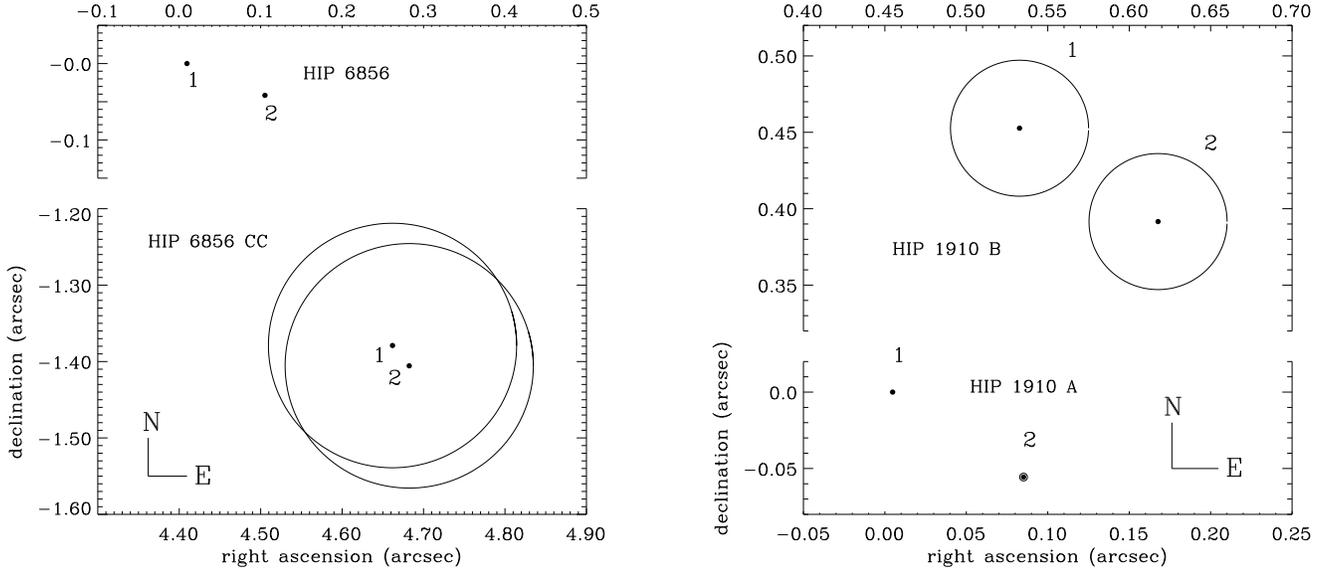

\centering
\hspace{0.2cm}
\includegraphics[width=8 cm, height=7.5cm]{h4174F3.epsi}
\hfill
\includegraphics[width=8 cm, height=7.5cm]{h4174F4.epsi}
\hspace{0.2cm}
\caption{\label{pm}{Left: We show the 1$\sigma$ error ellipses of the positions of the companion candidate to HIP~6856 relative to HIP~6856 on 12 November 2000 and 29 October 2001. We took into account the proper motion of HIP~6856, starting at ($\alpha,\delta)=(0,0)$ and then moving south-east to ($\alpha,\delta)=(0.100,-0.042)$. Regarding the ellipse errors, it is not possible to discriminate between bound or background objects for the HIP~6856 CC. Right: We show the 1$\sigma$ error ellipses of the position of the companion HIP~1910~B relative to HIP~1910~A on 12 November and 28 October 2001. We took into account proper motions of HIP~1910~A, starting at ($\alpha,\delta)=(0,0)$ which then moves south-east to ($\alpha,\delta)=(0.084,-0.056)$. The proper motions of HIP~1910~A and HIP~1910~B are significantly similar. The observed separations and P.A. are in total similar to $0.12\,\sigma$ in the case of a co-moving pair and $1.5\,\sigma$ discrepant from being a background object.}}
\end{figure*}

In the case of HIP~6856, the CC was detected on 12 November 2000 and re-observed on 29 October 2001. The corresponding astrometric results are reported in Table~\ref{tab:phot}. The proper motion of HIP~6856 reported in the Tycho catalog (H{\o}g et al. 2000) is $\mu_{\alpha}=104.8\pm1.2$~mas/yr and $\mu_{\delta}=-43.4\pm1.4$~mas/yr.
Taking the movement of HIP~6856 into account, we show positions of both visual components in Fig.~\ref{pm}, Left Panel. Because the primary was under the occulting mask the positional error ellipses are rather large. Therefore, these measurements do not conclusively determine the nature of the HIP~6856 CC as a background object or gravitationaly bound companion. 

The system HIP~1910~AB was also observed at two different epochs, 12 November 2000 and 28 October 2001. The astrometric measurements are presented in Table~\ref{tab:phot} and Fig.~\ref{pm}. Considering the proper motion of the star HIP~1910~A derived from the Tycho catalog (H{\o}g et al. 2000), $\mu_{\alpha}=88.0\pm1.8$~mas/yr and $\mu_{\delta}=-57.9\pm2.5$~mas/yr, we find that the proper motions of HIP~1910~A and B are significantly similar. The observed separations and P.A. are in total similar to $0.12\,\sigma$ in the case of a co-moving pair and $1.5\,\sigma$ discrepant from being a background object. Further spectroscopic measurements will allow us to confirm the nature of this companion. For the rest of this paper, we will regard HIP~1910~B as a companion to HIP~1910~A.

For HIP~6856 and HIP~1910, any possible orbital motion between the two epochs is negligible relative to the measurement errors. 
 
\subsection{Background object probability}

To help distinguish between background objects or gravitationaly bound companions one may estimate the probability for these CCs to be background objects. Based on the so-called ``Besan\c{c}on galactic model''  which is a synthetic model of the galaxy, the star density at a given galactic longitude and latitude can be obtained  as a function of magnitude in a given band. If we assume that the position of stars follow Poisson statistics, we can determine the probability $\eta_{F}$ to find at least one star with a magnitude in the K-band similar to a given CC within the field of view of the SHARPII camera ($12.7~\!''\times12.7~\!''$). The probability $\eta_{C}$ to find that star within a circle with a radius as large as the CC separation can also be estimated. The corresponding results for the four visual companions detected are presented in Table~2.

\begin{table}[h]
     \caption[]{Probabilities $\eta_{F}$ and $\eta_{C}$ for the stars to have respectively a background object with a magnitude in K-band similar to the one of their CC in the SHARPII camera field of view ($12.7~\!''\times12.7~\!''$) and within a circle with a radius as large as the CC separation.}
        \label{prob}
         \begin{tabular*}{\columnwidth}{@{\excs}lllll}
            \hline
            \hline
\noalign{\smallskip}
            Source     & \textit{l}.  &  \textit{b} &$\eta_{F}$   & $\eta_{C}$        \\
                       &  ($^o$)   &    ($^o$)              &($\%$) &($\%$)  \\
        \noalign{\smallskip}
            \hline      
 \noalign{\smallskip}
           HIP~108422 &  322.86      & -41.56      & $<1$  & $<0.1$ \\
           HIP~1910 &   308.43   & -54.65      & $<1$  & $<0.1$\\
          GSC~8047-0232 & 282.73 & -62.35    & $\sim2$ &  $\sim0.4$ \\
           HIP~6856 & 290.37 & -63.58     & $\sim5$  & $\sim2$ \\ 
        \noalign{\smallskip}
      \hline
   \end{tabular*}
 \end{table}

\section{HIP 108422}

\begin{figure*}[t]
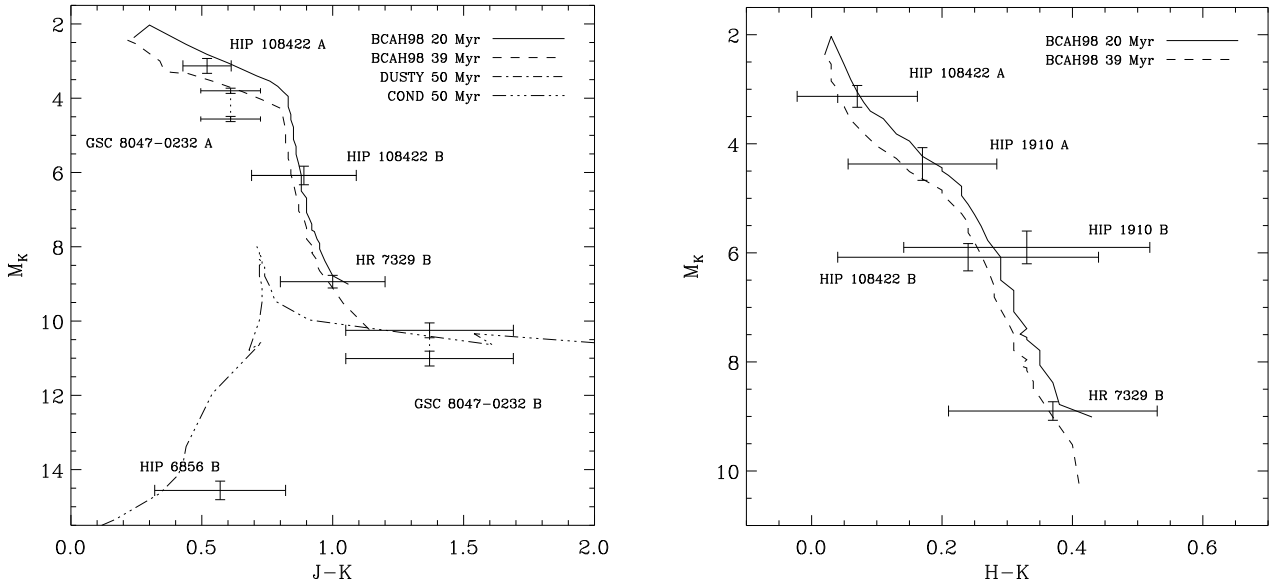

\centering
\hspace{0.5cm}
\includegraphics[width=8 cm]{h4174F5.epsi}
\hfill
\includegraphics[width=7.8 cm]{h4174F6.epsi}
\hspace{0.5cm}
\caption{\label{iso}{Left: CMD for the near-IR color ($J-K$) from the BCAH98 model and the CBAH00 DUSTY and COND models. The data of HIP~108422 and its CC, GSC~8047-0232 and its CC and the CC to HIP~6856 are overplotted with their uncertainties and also the sub-stellar companion HR~7329~B detected by Lowrance et al. (2000). In the case of GSC~8047-0232 and its CC, two distances are considered, 60~pc (lower) and 85~pc (upper). Right: CMD for the near-IR color ($H-K$) from the BCAH98 model. The data of HIP~108422 and its CC, and HIP~1910 A and B are overplotted.}}
\end{figure*}

When astrometry or spectroscopy is not available, a test to see if the CCs may be bound is to compare their photometry to evolutionary model predictions. We used the evolutionary tracks of Baraffe et al. (2002, 1998) based on a dust-free and non-gray atmosphere model (BCAH98). This model is an appropriate description for stellar objects with effective temperature $T_{\rm{eff}}>2300$K. The near-IR photometry of HIP~108422 (G8, $54.9^{+3.8}_{-3.3}$ pc) and its CC is compared to the color-magnitude diagram (CMD) for the near-IR color ($J-K$) based on an age of 40~Myr (see Fig.~\ref{iso}). The resulting stellar parameters of both objects are reported in Table~\ref{tab:stell1}. The model predictions for HIP~108422 are consistent with the G8 spectral type of the unresolved system HIP~108422 (Zuckerman et al. 2001a) which can be attributed to the primary star. The CC, detected at $1.97\pm0.02~\!''$, has colors similar to an M3-M5 dwarf.

\section{GSC 8047-0232}

No trigonometric parallaxes have been measured for the system GSC~8047-0232. Torres et al. (2000) initially supposed this star to be part of the Horologium association located at $60\pm25~$pc. Zuckerman et al. (2001a) suggest this system to be significantly further away from the large common Tucana-Horologium association, located within 50~pc from earth. To compare our data to model isochrones, we then consider two different distances: 60~pc and 85~pc. Figure~4 shows that BCAH98 and DUSTY CBAH00 models give a better fit for 85~pc than 60~pc, consistent with the greater hypothesized distance. We therefore adopt a distance of 85~pc to derive physical parameters of this possible binary.

The measured spectral type of the unresolved system GSC~8047-0232 is K3V (Torres et al. 2000) and can be attributed to the primary star. The color measurements ($J-K = 0.61 \pm 0.16$) and the stellar parameters deduced from the BCAH98 model isochrones (see Table~\ref{tab:stell1}) are in good agreement with the Kenyon \& Hartman (1995) color table. If we consider the case of the CC to GSC~8047-0232, we see in Fig.~4, that this object falls into the brown dwarf regime with a near-IR color ($J-K = 1.35 \pm 0.45$). As mentioned by Baraffe et al. (2002), the BCAH98 model is dust-free and not really appropriate to explain the near-IR colors of late M-dwarfs and L-dwarfs. As dust must be taken into account, we have used the CBAH00 DUSTY model to derive the physical parameters of the GSC~8047-0232 CC (see Table~\ref{tab:stell1}), based on the CMD for the near-IR color ($J-K$) and an age of 50~Myr (Fig.~4). The physical parameters of the GSC~8047-0232 CC are consistent with a sub-stellar object of 20 to 40~M$_{\rm{Jup}}$ and effective temperature between 1760K and 2475K.

The link between near-IR colors, effective temperature and spectral type is well constrained by the model predictions for late M dwarfs, but this is not really the case for the L dwarfs. The treatment of dust, particularly the grain condensation in the photosphere, is challenging and the $J-K$ color can vary $\sim0.5$ magnitude for L dwarfs of the same type (Leggett et al. 2002; Reid et al. 2001; Kirkpatrick et al. 2000). Leggett et al. (2002) have recently compared the infrared photometry of 58 late M, L and T dwarfs to the corresponding spectral classification done by Geballe et al. (2002). Based on these results, our photometry of the CC to GSC~8047-0232 is consistent with M7-L9 dwarfs. Clearly, new observations (without spike contamination) in photometry and spectroscopy are needed to determine precisely the stellar parameters and spectral type of this CC. If the sub-stellar nature of the object is confirmed, this would be the third brown dwarf companion known among young, nearby associations with TWA~5~B in the TW Hydrae association (Lowrance et al. 1999; Neuh\"auser et al. 2000) and HR~7329~B in the $\beta$ Pictoris group (Lowrance et al. 2000; Guenther et al. 2001; Zuckerman et al. 2001b) being the other two.

\begin{table}[h]
     \caption[]{Stellar parameters of HIP~108422 and its CC are derived from the BCAH98 model for an age of 40~Myr. This model is also used for the star GSC~8047-0232, but the CBAH00 DUSTY model is used for its CC for an age of 50~Myr. }
        \label{tab:stell1}
         \begin{tabular*}{\columnwidth}{@{\excs}llll}
            \hline        
\noalign{\smallskip}
            Source     & M              &  T$_{\rm{eff}}$         &L             \\
                       & (M$_{\odot}$)    &  (K)                 &(L$_{\odot}$)  \\
        \noalign{\smallskip}
            \hline      
 \noalign{\smallskip}
           HIP\,108422 & [0.9, 1.15]       & [4760,5638]      &[0.5,0.8]     \\
           HIP\,108422\,CC& [0.2, 0.35]      & [3301,3455]      &[0.013,0.03]     \\
        \noalign{\smallskip}
            \hline
\noalign{\smallskip}
            GSC\,8047-0232 & [0.8, 0.9]      & [4203,4764]               &[0.02,0.45]     \\
        \noalign{\smallskip}
            \hline
\noalign{\smallskip}
\noalign{\smallskip}
\noalign{\smallskip}
\hline
\noalign{\smallskip}
            GSC\,8047-0232\,CC & [0.02, 0.04]      & [1760, 2475]      &[2e-4,8e-4]     \\
\noalign{\smallskip}
\hline
\end{tabular*}
\end{table}

\section{HIP 6856}

In the close vicinity of HIP~6856 (K1V, $37.1_{-1.6}^{+1.8}~$pc), the very faint CC detected with the coronagraphic mode of ADONIS/SHARPII has a color ($J-K = 0.57 \pm 0.35$) (see Table~\ref{tab:phot} and Fig.~4). It would have an absolute magnitude $M_{K}=14.56$ if it was a companion. The photometry is not consistent with the CBAH00 DUSTY model predictions, demonstrating that this object is unlikely to be a late M dwarf or L dwarf. However, the photometry is  consistent with cooler objects such as T dwarfs, typically with $T_{\rm{eff}}<$~1300K and where the CO bands are replaced by strong methane absorption in H and K bands responsible for the weakening of the $J-K$ color (Geballe et al. 2002). This is poorly modelled in both the extreme CBAH00 COND and DUSTY models. Dahn et al. (2002) and Leggett et al. (2002) have presented near-IR measurements of T dwarfs that are consistent with the photometry found for the CC to HIP~6856. Our present astrometric results do not distinguish whether the object is bound or not to HIP~6856 and future astrometric or spectroscopic observations are necessary to confirm its nature. \textbf{If the CC is a T-type companion, its mass is $\sim6-8$~M$_{\rm{Jup}}$ and at 37.1~pc its angular separation corresponds to a projected physical separation of $\sim180$\,AU. }

\section{HIP~1910~A and B}

AO imaging allows us to resolve the source HIP~1910 (M0V, $46.3_{-4.3}^{+5.3}$~pc) as a close binary, and photometry was obtained in $H$ and $K$. Based on the BCAH98 CMD for the color ($H-K$) for an age of 40~Myr (see Fig.~4, Right Panel), we find that the stellar parameters of HIP~1910~A are consistent with a M0V star (contribution of HIP~1910~B negligible). HIP~1910~B, detected within the errors at the same separation and position angle from HIP~1910~A, is consistent with being a companion. It has photometry consistent with an M3-M5 star. Derived stellar parameters of both components are reported in Table~\ref{stell2}.

\begin{table}[h]
     \caption[]{Stellar parameters of HIP~1910~A and B derived from the BCAH98 model for an age of 40~Myr.}
        \label{stell2}
          \begin{tabular*}{\columnwidth}{@{\excs}llll}  
            \hline
            \hline
\noalign{\smallskip}
            Source     & M              &  T$_{\rm{eff}}$         &L             \\
                       & (M$_{\odot}$)    &  (K)                 &(L$_{\odot}$)  \\
        \noalign{\smallskip}
            \hline      
 \noalign{\smallskip}
           HIP\,1910\,A & [0.62,0.8]       & [3710,4203]      &[0.08,0.2]     \\
  HIP\,1910\,B &  [0.2, 0.4]      & [3360, 3495]      & [0.02,0.04]   \\
        \noalign{\smallskip}
      \hline
   \end{tabular*}
\end{table}

\section{Conclusion}

We present results of an AO imaging survey of the common associations of Tucana and Horologium. Four candidate companions (CCs) were detected in a sample of 24 stars. We first estimated the probability for these objects to be background objects. We then compare predictions of low-mass stellar and sub-stellar evolutionary models to our photometry to test if the CCs might be physically bound objects. As already mentioned, these models provide an accurate comparison with observations of such young objects but are still not as reliable at an age of a few tens Myr as they depend on the initial conditions. We find that the HIP~108422 CC is likely to be an M3-M5 dwarf, to be confirmed by spectroscopy. Based on the Chabrier et al. (2000) DUSTY model, the CC to GSC~8047-0232 falls into the brown dwarf regime with a deduced mass of 20 to 40~M$_{\rm{Jup}}$. In comparison to late-M, L and T dwarf observations (Leggett et al. 2002; Geballe et al. 2002), this object could be an M7-L9 dwarf. If confirmed, this object would be the third brown dwarf companion detected among young, nearby associations along with TWA~5~B in the TW Hydrae association (Lowrance et al. 1999; Neuh\"auser et al. 2000) and HR~7329~B in the $\beta$ Pictoris moving group (Lowrance et al. 2000; Guenther et al. 2001; Zuckerman et al. 2001b). 

The photometry of the CC detected in the close vicinity of HIP~6856 is consistent with cooler objects like T dwarfs. Our astrometry did not enable us to conclude whether the CC object is a true companion of HIP~6856. Finally, we demonstrate that HIP~1910~B has a common proper motion with HIP~1910~A and that further spectroscopic measurements are needed to confirm this companionship. Based on comparison of the photometry with model predictions, we also show that HIP~1910~B may be an M3-M5 dwarf.

\begin{acknowledgements}
We would like to thank Eric Becklin for his contribution to this AO imaging campaign on November 2000 and on October 2001. We would like to thank also Isabelle Baraffe who kindly sent us the COND model predictions reported in Fig.~4 for comparison to our data and Franck Marchis who helped us for the ADONIS/SHARPII astrometric calibrations. We thank the staff of the ESO 3.6m telescope and particulary Kate Brooks for her outstanding support and finally Ralph Neuh\"auser for his precious comments in acknowledgements.    
\end{acknowledgements}




\begin{thebibliography}{twocolumn}


\bibitem[1998]{bara98} Baraffe I., Chabrier, G., Allard, F. \& Hauschildt, P.H ,1998, A\&A, 337, 403
\bibitem[2002]{bara02} Baraffe, I., Chabrier, G., Allard, F. \& Hauschildt, P.H. 2002, A\&A, 382, 563

\bibitem[1997]{burr} Burrows, A., Marley, M., Hubbard, W.B. et al. 1997, ApJ, 491, 856

\bibitem[2000]{chab} Chabrier, G., Baraffe, I., Allard, F. \& Hauschildt, P.H. 2000, ApJ, 542, 464

\bibitem[2000]{cona} Conan, J.-M., Fusco, T., Mugnier, L. et al. 2000, Proc. SPIE, Vol. 4007, ed. P.L. Wizinowich, p.913

\bibitem[1998]{dahn}Dahn, C. C., Harris, H. C., Vrba, F. J. et al. 2002, AJ, 124, 1170

\bibitem[1999]{fusco1}Fusco, T., V\'eran, J.-P., Conan, J.-M. \& Mugnier, L. 1999, A\&AS, 134, 193
 
\bibitem[1998]{geb} Geballe T. R., Knapp, G. R., Legget, S. K. et al. 2002, ApJ, 564, 466

\bibitem[2000]{guen}Guenther, E.W., Neuh\"auser, R., Hu\'elamo, N., Brandner, W. \& Alves, J. 2001, A\&A, 365, 514

\bibitem[2000]{hog}H{\o}g, E., Fabricius, C., Makarov, V.V. et al. 2000, A\&A, 355, 27  

\bibitem[1997]{kast} Kastner, J. H., Zuckerman, B., Weintraub, D. A. \& Forveille, T. 1997, Science, 277,  67 

\bibitem[1995]{kenyon}   Kenyon, S. J. \& Hartmann, L. 1995, ApJS, 101, 117
%
\bibitem[2000]{kirk00} Kirkpatrick, J.D., Reid, I. N., Liebert, J. et al. 2000, AJ, 120, 447
%
\bibitem[2002]{leg}Legget, S. K., Fan, X., Golimowski, D. A. et al. 2002, ApJ, 564, 452

\bibitem[1999]{lowr99} Lowrance, P. J., McCarthy, C., Becklin, E. E. et al. 1999, ApJ, 512, L69

\bibitem[2000]{lowr} Lowrance, P. ,Schneider, G., Kirkpatrick, J. D. et al. 2000, ApJ, 541, 390

\bibitem[2000]{neu} Neuh\"auser, R., Guenther, E.W., Petr, M.G. et al. 2000, A\&A, 360, L39

\bibitem[2001]{rei} Reid, I.N., Burgasser, A.J., Cruz, K.L., Kirkpatrick, J. D. \& Gizis, J.E. 2001, AJ, 121, 1710

\bibitem[2000]{ste} Stelzer, B. \& Neuh\"auser, R. 2000, A\&A, 361, 581

\bibitem[2000]{torr} Torres, C.A.O., Da Silva, L., Quast, G.R., de la Reza, R. \& Jilinski, E. 2000, AJ, 120, 1410

\bibitem[1998]{veran98} V\'eran, J.-P. \& Rigaut, F. 1998, SPIE, 3353, 426

\bibitem[2000]{zuck}Zuckerman, B. \& Webb, R.A. 2000, ApJ, 535, 959

\bibitem[2001a]{zucka} Zuckerman, B., Song, I. \& Webb, R.A. 2001a, ApJ, 559, 388 
\bibitem[2001b]{zuckb} Zuckerman, B., Song, I., Bessel, M.S. \& Webb, R.A.  2001b, 562, L87





\end{thebibliography}
\end{document}